%% file: main.tex
\documentclass[
    twocolumn,
	prd,
	amssymb,
	preprintnumbers,superscriptaddress,
	nofootinbib]{revtex4-1}

\pdfoutput=1
\usepackage{paralist}
\usepackage{graphicx}
\usepackage{enumitem}
\usepackage{latexsym}
\usepackage{amsfonts}
\usepackage{amssymb}
\usepackage{xcolor}
\usepackage[export]{adjustbox}
\usepackage{amsmath}
\usepackage[thinlines]{easytable}
\usepackage{slashed}
\usepackage{dcolumn}
\usepackage{verbatim}
\usepackage{float}
\usepackage{multirow}
\usepackage{xspace}
\usepackage[normalem]{ulem}
\usepackage[
pdfauthor={Djuna Croon}]{hyperref}
\usepackage{tabularx}
\usepackage{lettrine}
\usepackage{setspace}

\input Zallman.fd

\LettrineTextFont{\itshape}

\input{universalnewcommands.tex}

\newcommand{\sign}{\text{sign}}

\allowdisplaybreaks

\begin{document}

\title{Supernova microlensing as a probe of ultracompact minihalos and primordial cosmology}

\author{Djuna Croon} \email{djuna.l.croon@durham.ac.uk}
\affiliation{Institute for Particle Physics Phenomenology, Department of Physics, Durham University, Durham DH1 3LE, U.K.}
\author{Sergio Sevillano Mu\~{n}oz} \email{sergiosm@sas.upenn.edu}
\affiliation{Center for Particle Cosmology, Department of Physics and Astronomy, University of Pennsylvania, Philadelphia, Pennsylvania 19104, USA}
\author{Miguel Zumalacarregui} \email{miguel.zumalacarregui@aei.mpg.de}
\affiliation{Max Planck Institute for Gravitational Physics (Albert Einstein Institute) \\
 Am Mühlenberg 1, D-14476 Potsdam-Golm, Germany}
\affiliation{Instituto de Física Teórica UAM/CSIC
C/ Nicolás Cabrera 13-15 
Universidad Autónoma de Madrid
Cantoblanco, Madrid 28049, Spain}

\date{\today}

\begin{abstract}
We propose using supernova microlensing to search for ultracompact minihalos and other extended dark structures seeded by enhanced primordial small-scale power. Unlike conventional microlensing searches, which are often optimized for primordial black holes, cosmological supernovae are sensitive to lenses with physical sizes comparable to their Einstein radii (up to $\sim {\rm pc}\left(M/10^4M_\odot\right)^{1/2}$ scale), opening a qualitatively different region of dark object parameter space. We compute the extended object detection efficiency relative to point lenses. We then show how projected limits on the abundance of ultracompact minihalos can be mapped, in benchmark formation scenarios, onto constraints on the primordial power spectrum. With the Vera C. Rubin Observatory expected to deliver an unprecedented LSST supernova sample, supernova microlensing is poised to become a timely and complementary probe of small-scale structure and the physics of the early Universe.
\end{abstract}

\preprint{IPPP/26/61, IFT-UAM/CSIC-26-123}

\maketitle

\section{Introduction}
The small-scale primordial power spectrum remains one of the least directly tested predictions of early-Universe cosmology. Enhanced power on these scales can lead to the formation of compact dark objects, including primordial black holes (PBHs), ultracompact minihalos (UCMHs), and other dense dark matter (DM) substructures \cite{Ricotti:2009bs}. Detecting or constraining such objects via their gravitational signals would provide a direct window onto inflation and early Universe dynamics \cite{Bringmann:2025cht}. 

Gravitational microlensing is one of the cleanest probes of dark objects. Its sensitivity depends on the size of the lens relative to the physical Einstein radius $R_E$. If the mass of an extended dark object (EDO) is spread over radii comparable to or larger than \(R_E\), the magnification profile and effective lensing cross-section can be strongly modified \cite{Croon:2020wpr,Bai:2020jfm,Croon:2020ouk,CrispimRomao:2024nbr,Croon:2025yfj}.
This finite-lens suppression implies conventional Galactic microlensing is not sensitive to UCMH \cite{Bringmann:2025cht} (see \cite{Delos:2023fpm} for an exception). However, they can be constrained from other gravitational probes such as via their effect on the CMB due to accretion~\cite{Croon:2024rmw}, on wide-binary evaporation using GAIA~\cite{Ramirez:2022mys} and ultra-faint dwarf galaxies~\cite{Graham:2023unf,Olea-Romacho:2026pgn}, and gravitational waves~\cite{Fairbairn:2022xln,Tambalo:2022wlm,Choi:2026lsa}. Future probes will also be able to place constraints on UCMHs by using astrometry~\cite{VanTilburg:2018ykj,Mishra-Sharma:2020ynk,VanTilburg:2026eym} and pulsar timing arrays~\cite{Baghram:2011is, Dror:2019twh, Ramani:2020hdo}.

In this work we propose supernova microlensing as a probe of EDOs, including UCMH, and, by extension, of the primordial small-scale power spectrum. Supernovae are particularly well suited to this purpose because they are bright sources which can be observed at cosmological distances. Compact objects along the line of sight modify the magnification distribution of standardizable candles. A population of DM lenses produces a non-Gaussian distribution of Hubble residuals: most lines of sight are slightly demagnified relative to a homogeneous universe, while a small fraction are significantly magnified by close lensing encounters. The absence of an excess high-magnification tail in the observed SN Hubble diagram can therefore be used to constrain the fraction of matter in compact objects \cite{1991ApJ...374...83R,Metcalf:1999qb,1999A&A...351L..10S,Bosca:2022viy,Zumalacarregui:2017qqd}. Most recently, \cite{DES:2024ffp} applied this method to the Dark Energy Survey five-year supernova sample. 

The large SN distances imply large physical Einstein radii, allowing sensitivity to objects much more spatially extended than those probed by conventional Galactic microlensing. For a lens of mass $M$, the physical Einstein radius in the lens plane is
\begin{equation}
\label{eq:RE}
\begin{split}
    R_E &=
\left(
\frac{4GM}{c^2}
\frac{D_l D_{ls}}{D_s}
\right)^{1/2} \\
&\simeq
2.8\times 10^3~{\rm AU}
\left(\frac{M}{M_\odot}\right)^{1/2}
\left(
\frac{D_l D_{ls}/D_s}{1~{\rm Gpc}}
\right)^{1/2}.
\end{split}
\end{equation}
where $D_l$, $D_s$, and $D_{ls}$ are angular-diameter distances to the lens, to the source, and between the lens and source, respectively, and where on the second line we have assumed $D_l D_{ls}/D_s \sim 1~{\rm Gpc}$ for a representative cosmological geometry.
This size range is well matched to the expected scales of dense DM substructures formed from enhanced primordial perturbations \cite{Bringmann:2025cht}. UCMHs are a particularly interesting target because their abundance is directly related, within a specified formation model, to the amplitude of primordial density fluctuations on scales far smaller than those probed by the cosmic microwave background. Supernova microlensing therefore offers a route from transient observations to constraints on early-Universe physics where Galactic microlensing is limited.

We begin in Section~\ref{sec:lens effects} by introducing the lensing equations and efficiencies of extended lenses for point-like and extended sources. Based on this, in Section~\ref{sec:EDOconstraints} we demonstrate how constraints on extended objects can be derived. Then, in Section~\ref{sec:powerspectrum}, we translate the resulting UCMH limits into constraints on the primordial curvature power spectrum. Finally, we conclude in Section~\ref{sec:conclusion}.

\section{Finite lens effects}\label{sec:lens effects}
We will now discuss how SNe constraints on compact objects generalize to extended lenses, first in the point-source approximation and then for finite source radius.

\subsection{Point source}
To quantify the loss of sensitivity when the lens is spatially extended, we compare the lensing cross-section of an UCMH-like density profile to that of a point lens with the same total mass. For that, we follow a similar formalism as in Refs.~\cite{Croon:2020wpr}.

We start by considering a radially symmetric lens of density $\rho(r)$ with total mass $M=4\pi\int^\infty_0drr^2\rho(r)$. For the power law profiles considered below we impose a finite outer radius $r_m$, as in Ref.~\cite{Croon:2020ouk}, and parameterise the physical extent of the lens by $r_{90}$, the radius enclosing $90\%$ of its total mass.
The true angular separation between the lens center and the source, $\beta$, is related to the observed angular separation between the lens center and the image, $\theta$, by the lens equation
\begin{equation}\label{eq: beta}
    \beta(\theta)=\theta-\frac{\theta_E^2}{\theta}\frac{M(\theta)}{M},
\end{equation}
where $M(\theta)$ is the projected mass enclosed within angular radius $\theta$ on the lens plane. It is given by
\begin{equation}
\begin{split}
    M(\theta)
    &=2\pi D_l^2\int_0^\theta d\theta'\,\theta'\,\Sigma(\theta'),\\
    \Sigma(\theta)
    &=\int_{-\infty}^{\infty} dz\,
    \rho\left(\sqrt{D_l^2\theta^2+z^2}\right),
\end{split}
\end{equation}
where $\Sigma(\theta)$ is the surface mass density projected onto the lens plane and $D_l$ is the angular-diameter distance to the lens. The relevant dimensionless parameter characterizing the size of an extended lens is $r_{90}/R_E$, where $R_E$ is the Einstein radius of a point lens with the same total mass and lens-source geometry, given in \eqref{eq:RE}.
From this, we can also define the Einstein angle as the angular size subtended by the Einstein radius in the lens plane, $\theta_E=R_E/D_l$.

For a given microlensing event, its magnification is defined as the ratio of the observed image flux to the unlensed source flux. This is given for each image, $\theta_i$, by
\begin{equation}\label{eq: mu}
\mu_i(\theta_i)=\left|\frac{\theta}{\beta}\frac{d\theta}{d\beta}\right|_{\theta\to\theta_i},
\end{equation}
where the total magnification $\mu$ is given by the sum of the individual $\mu_i$. For a point-like object, the magnification can be analytically solved by
\begin{equation}
\mu_{\rm pt}(u)
=
\frac{u^2+2}{u\sqrt{u^2+4}},
\end{equation}
where $u=\beta/\theta_E$ is the impact parameter in units of Einstein radii. For an extended lens this is more complicated, as $\beta(\theta)$ also depends on the object's size, $r_{90}$ through $M(\theta)$. Therefore, for a given detection threshold, specified as a minimum magnification $\mu_{\rm th}$, we numerically compute the largest impact parameter $u_{\rm th}$ for which the lens produces an observable signal
\begin{equation}
\mu(u_{\rm th}; {r_{90}}/{R_E} ) = \mu_{\rm th}.
\end{equation}
 
We can then define the relative efficiency of the extended lens as the ratio of the detectable lensing cross-section to that of a point lens,
\begin{equation}
\epsilon({r_{90}}/{R_E};\mu_{\rm th})
\equiv
\frac{\sigma_{\rm ext}({r_{90}}/{R_E};\mu_{\rm th})}
{\sigma_{\rm pt}(\mu_{\rm th})}
=
\left[
\frac{u_{\rm th}^{\rm ext}}
{u_{\rm th}^{\rm pt}}
\right]^2 .
\label{eq:efficiency}
\end{equation}
This definition isolates the change in detectability caused by the finite spatial extent of the lens, while holding the total lens mass fixed.
The square appears because the relevant quantity for a randomly distributed population of lenses is the detectable area, or lensing cross-section, which scales as $\sigma \propto b_{\rm th}^2 = (u_{\rm th}R_E)^2$, rather than linearly with the maximum detectable impact parameter.
This differs from the convention commonly used for Galactic microlensing of moving stellar sources, where the event rate is proportional to the linear width swept out by the lens-source trajectory rather than to the instantaneous cross-sectional area~\cite{Croon:2020wpr,Croon:2020ouk,CrispimRomao:2024nbr}. In that convention, the factor $u_{\rm th}^{\rm ext}/u_{\rm th}^{\rm pt}$ may be absorbed into an effective Einstein radius, whereas here the relevant efficiency is the square of this ratio.

\begin{figure}
    \centering
    \includegraphics[width=0.9\linewidth]{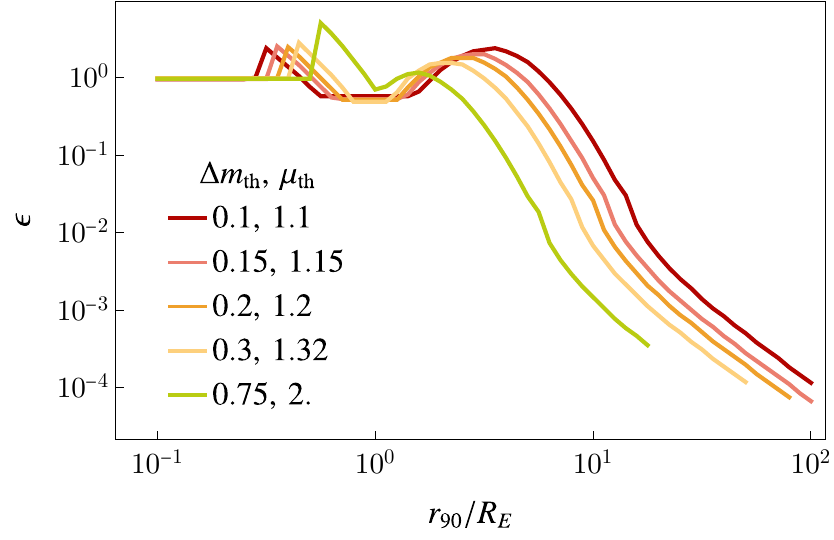}
    \caption{Efficiency as calculated in \eqref{eq:efficiency} for lenses with profile $\rho \propto r^{-3/2}$. Here $ \mu_{\rm th} =2$ corresponds to the threshold used in \cite{Zumalacarregui:2017qqd}. 
  We recover the point-like limit $\epsilon \to 1$ for $r_{90} \ll R_E$ and diffuse limit $\epsilon \to 0$ for $r_{90} \gg R_E$ as expected. Intermediate lens sizes can feature $\epsilon >1$ as a result of the finite size of the caustics of extended lenses of this profile.
    }
    \label{fig:efficiency}
\end{figure}

In practice we express the detection threshold as a magnitude brightening,
\begin{equation}
|\Delta m_{\rm th}|
=
2.5 \log_{10}\mu_{\rm th},
\end{equation}
and evaluate $\epsilon$ for several representative thresholds,
\begin{equation}
|\Delta m_{\rm th}|
= 0.1,\ 0.15,\ 0.2,\ 0.3, \ 0.75,
\end{equation}
which span optimistic population-level sensitivity to more conservative single-object perturbations. We show the results in Fig.~\ref{fig:efficiency} for an object with $\rho \propto r^{-3/2}$.
As expected, for compact minihalos, $r_{90}\ll R_E$, the lens behaves approximately as a point mass and we find $\epsilon\simeq 1$. 
As the UCMH radius becomes comparable to the Einstein radius, finite-size effects modify the magnification profile and the efficiency can deviate from unity. In a range near $r_{90}/R_E = 1$, the extended profile increases the impact parameter range over which a fixed threshold magnification is reached, leading to $\epsilon>1$. This enhancement is associated with the caustic structure of the finite lens: unlike a point lens, for which the image multiplicity is fixed for all non-zero impact parameters, an extended mass distribution can undergo changes in the number of images as the source crosses caustics. The resulting localized magnification enhancements increase the effective cross-section for exceeding a fixed magnification threshold. 
For sufficiently extended objects, $r_{90}\gg R_E$, the projected surface density is too diffuse to generate a strong microlensing perturbation, and the efficiency rapidly falls below the point-lens value.

\subsection{Extended source}
\label{sec:finitesource}
In reality, supernovae are not static point sources. Their photospheric radii evolve with time, introducing finite-source effects that can smooth or reshape the microlensing signal. 
Finite-source effects become important when the physical size of the supernova photosphere is comparable to or larger than the Einstein radius, setting the low-mass scale below which the point-lens approximation loses sensitivity.

A type Ia supernova reaches peak brightness roughly $t_{\rm peak}\sim20\,{\rm days}$ after explosion, with a characteristic ejecta velocity of order $v_{\rm ej}\sim10^{4}\,{\rm km\,s^{-1}}$ (e.g. \cite{Riess:1999th,Hillebrandt:2000ga}).  The characteristic photospheric radius at peak is therefore 
$
  R_S \sim v_{\rm ej}\,t_{\rm peak}
  \approx 115\,{\rm AU}.
$ 
We define $r_S=R_sD_l/D_s$ as the source radius projected onto the lens plane.
Equating $r_S = R_E$ yields
the mass at which the photosphere just fills the Einstein radius,
\begin{equation}
\begin{split}
  M_{\rm ps} &= \frac{c^2 R_S^2}{4G}\,\frac{D_L}{D_S D_{LS}} \\
  &\approx 1.6\times10^{-3}\,M_\odot
  \left(\frac{R_S}{115\,{\rm AU}}\right)^{2}
  \left(\frac{D_S D_{LS}/D_L}{1\,{\rm Gpc}}\right)^{-1}.   
\end{split}
\label{eq:Mfinitesource}
\end{equation}
For $M\lesssim M_{\rm ps}$ the source can no longer be treated as point-like.

In this subsection we compute the interplay between finite source and finite lens effects. To do so, in units of $R_E$, we parametrize the boundary of the source as $\bar{u}(\varphi)=\sqrt{u^2+(r_S/R_E)^2+2u (r_S/R_E) \cos\varphi}$, where $u$ is the distance between the center of the source and that of the lens. The lensing equation for a given infinitesimal $\varphi$ is \cite{1994ApJ...430..505W,Croon:2020ouk}
\begin{equation}
\label{eq:lensing-extended}
    \bar{u}(\varphi)=\tau(\varphi)-\frac{1}{\tau(\varphi)}\frac{M(\tau(\varphi))}{M},
\end{equation}
from which we obtain the positions of images at $\tau_i(\bar{u}(\varphi))$, with $i$ labelling the different solutions.

\begin{figure}[t]
    \centering
    \includegraphics[width=0.9\linewidth]{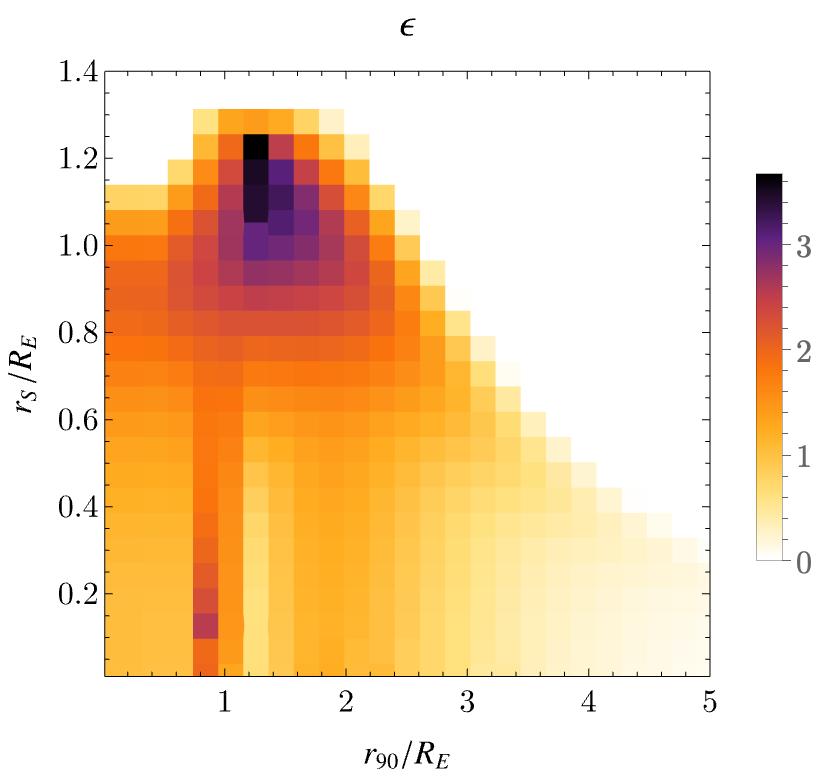}
    \caption{Efficiency as calculated in \eqref{eq:efficiency} with lens profile $\rho \propto r^{-3/2}$ and $ \mu_{\rm th} =2$. For finite lenses, the extended source can increase the efficiency beyond the point-lens limit set by $\mu_{\rm max}$ ($r_S/R_E<1.15$ here) because part of the source overlaps with the caustics of the extended lens.
    }
    \label{fig:efficiencyFS}
\end{figure}

Once the solutions are known, we can calculate the magnification by comparing the area of the lensed to the unlensed image, yielding
\begin{equation}
    \mu_i=\eta_i\frac{R_E^2}{\pi r_S^2}\int^{2\pi}_{0}d\varphi \,\frac{1}{2}\tau_i^2(\varphi),
\end{equation}
where $\eta_i\equiv \sign(d\tau_i^2/d\bar{u}^2|_{\varphi\to\pi})$ is the parity of the image. The total magnification can now be obtained by summing over each image, $\mu_i$. As for the point source case, the lensing equation can be solved analytically for a point-like lens, finding  $|\tau_\pm|=|\bar{u}|/2\times|1\pm\sqrt{1+4/\bar{u}^2}|$, while for an extended lens we must solve this numerically. We can see that the magnification saturates at $\mu_{\rm max}\simeq\sqrt{1+4(R_E/r_S)^2}$ and the lensing signal is progressively suppressed, which defines the
low-mass boundary of our sensitivity. In Fig.~\ref{fig:efficiencyFS} we demonstrate that this conclusion continues to hold for extended lenses, where we used the definition from Eq.~\eqref{eq:efficiency}.

The efficiency map exhibits a clear asymmetry between the two size parameters. This can be understood as follows. A finite source imposes a hard ceiling on the achievable magnification: a source of radius $r_S$ centred on the lens cannot be magnified beyond $\mu_{\max}$, which falls monotonically to unity with increasing $r_S/R_E$ by conservation of surface brightness. A threshold $\mu_{\rm th}$ therefore excludes \emph{all} configurations with $r_S/R_E>2/\sqrt{\mu_{\rm th}^2-1}$ ($\simeq1.15$ for $\mu_{\rm th}=2$ and point-like lens), independently of the impact parameter. This sets the sharp upper boundary of the efficient region, and is the same effect that fixes the low-mass floor of SN microlensing constraints \cite{Zumalacarregui:2017qqd,DES:2024ffp}. A finite lens carries no such ceiling: since the lensing observable is the projected enclosed mass $M(<\theta)$, an extended lens behaves almost as a point mass provided most of its mass lies within $\sim R_E$; spreading the profile dilutes the central enclosed mass rather than capping the magnification.

\section{Constraints on extended dark objects}\label{sec:EDOconstraints}

The relevant observable in supernova microlensing is its cumulative effect on the population-level magnitude distribution. In practice, this distribution receives contributions both from compact-object microlensing and from lensing by large-scale structure. The latter produces a redshift-dependent magnification scatter and skewness, sourced by the projected matter density along the line of sight. This contribution is present even in the absence of compact objects and must therefore be included as a background lensing distribution, or marginalised over, when extracting limits on compact or extended dark lenses. Compact objects are distinguished by the sharper high-magnification tail associated with rare close encounters, while large-scale structure produces a smoother, more continuous contribution to the residual distribution. The full analysis of these two effects on supernova microlensing was originally done in Ref.~\cite{Zumalacarregui:2017qqd} for point-like objects. Given this prescription, we will recast the results of \cite{DES:2024ffp}, for extended objects in this section.

The finite size of the supernova relative to a point lens is encoded in an \emph{effective lens fraction}: a lens at redshift $z_L$ contributes to the constraint only if it can magnify the source above a threshold, so the true compact-object fraction is replaced in the likelihood by $\tilde\alpha(z_S,M)=\alpha\,f_L(z_S,M)$ with
\begin{equation}\label{eq: fL}
    f_L
    =\frac{1}{\tau(z_S)}\int_0^{z_S}\frac{d\tau}{dz_L}\,
    \Theta\!\big(\mu_{\rm max}(z_L,z_S,M)-\mu_{\rm th}\big)\,dz_L .
\end{equation}
The differential optical depth $d\tau/dz_L\propto\alpha\Omega_M H_0^2\,(1+z_L)^2 D_L D_{LS}/(H(z_L)D_S)$ sets how much each lens redshift contributes and depends only on the \emph{total} mass in lenses and on the line-of-sight geometry. The dependence on an individual lens enters separately, through the selection function $\Theta(\mu_{\rm max}(z_L,z_S,M)-\mu_{\rm th})$: a lens is counted only if it can magnify the finite source above threshold, and $\mu_{\rm max}$ depends on the lens mass through its Einstein radius. 
We adapt this prescription for a spatially extended lens in the following way. We replace the binary selection $\Theta$ by the smooth per-lens efficiency $\epsilon({r_{90}}/{R_E};\mu_{\rm th})$: each lens retains a detectable cross-section modified by its finite extent rather than being counted or discarded. The effective fraction then becomes
\begin{equation}\label{eq: alpha eps}
    \begin{split}
        \tilde\alpha(z_S,M)=\alpha\,\langle\epsilon\rangle_\tau, \\
        \langle\epsilon\rangle_\tau\equiv\frac{1}{\tau(z_S)}\int_0^{z_S}
    \frac{d\tau}{dz_L}\,\epsilon\big({r_{90}}/{R_E}(z_L);\mu_{\rm th}\big)\,dz_L ,
    \end{split}
\end{equation}
the optical-depth-weighted average of the efficiency along the line of
sight, in direct analogy to Eq.~\eqref{eq: fL}.
To represent the DES sample, we use effective survey redshift $z_S\simeq0.47$ \cite{DES:2024ffp} in a flat $\Lambda$CDM cosmology ($\Omega_M=0.3$ and $H_0 = 67.4 \, \rm km/s/Mpc$). 
In contrast, $z_S\simeq 1$ is more representative of the LSST supernova sample.

We conservatively neglect any strengthening of the bound in the $\epsilon>1$ caustic region, the regime most sensitive to both the assumed source/lens distance and the finite-source smoothing of the supernova photosphere. 
Then, as in Ref.~\cite{Zumalacarregui:2017qqd}, we conservatively adopt a threshold $\mu_{\rm th} = 2$. This is conservative in two senses: it lies beyond the magnification of any observed supernova, and it requires sampling the inner lens region ($u_{\rm th}<1$ such that $\beta < \theta_E$) where extended profiles are most suppressed (see Fig.~\ref{fig:efficiency}).

The DES limit is formally a constraint on $\alpha < 0.12$, the fraction of the \emph{total} matter density in compact objects, rather than on $f_{\rm DM}$ directly~\cite{DES:2024ffp}.
The two normalisations differ only by the matter budget, $f_{\rm DM, pl}^{\rm DES} = \alpha\, \Omega_{\rm m}/\Omega_{\rm DM} \approx 1.19\,\alpha = 0.14$.
\begin{figure*}
    \centering
    \includegraphics[width=1\linewidth]{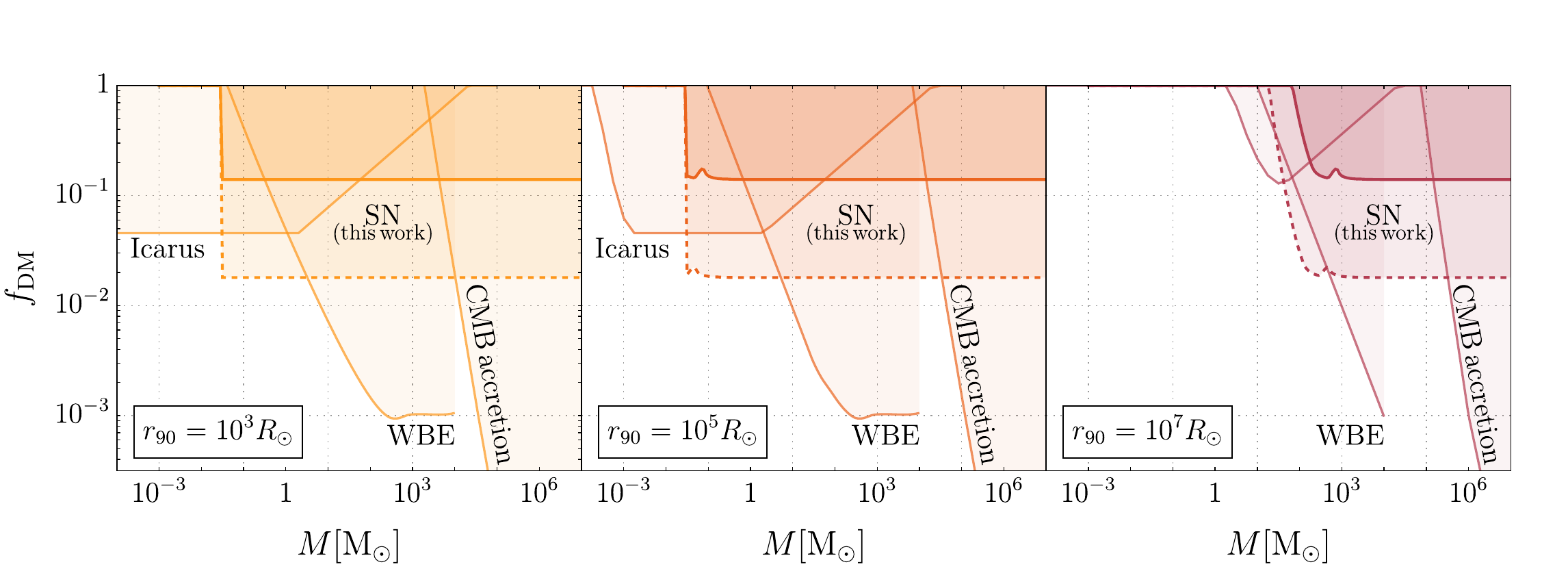}
    \caption{Dominant constraints in the large-mass limit for monochromatic EDOs with $\rho\propto r^{-3/2}$ and different radii, $r_{90}$. Supernova bounds are only effective for objects with $r_{90}<R_E$, yielding $f_{\rm DM}<0.14$ for DES (solid line) and $f_{\rm DM}<0.018$ for LSST as a conservative projection (dashed line). These bounds are compared with wide-binary evaporation bounds (WBE) from ultra-faint dwarf galaxies~\cite{Olea-Romacho:2026pgn}, caustic-crossing constraints from Icarus~\cite{Bringmann:2025cht,Croon:2025yfj}, and bounds from accretion~\cite{Croon:2024rmw}. Comparisons across a wider range of radii, set of constraints, and mass functions can be found in the \texttt{EDObounds} repository~\cite{Croon:2024jhd,Sevi2024EDObounds}.}
    \label{fig:edo_constraints}
\end{figure*}

The LSST projection is obtained by rescaling the DES dark object limit with the square root of the sample size,
\begin{equation}
    f_{\rm DM,pl}^{\rm LSST}
    =
    0.14
    \left(\frac{1532}{10^5}\right)^{1/2}
    \simeq 0.018,
\end{equation}
assuming a sample of $\sim 10^5$ usable SNe~Ia. This $N_{\rm SN}^{-1/2}$ scaling should be regarded as a conservative, background-limited forecast rather than the optimal scaling of a pure rare-event search. Indeed, a simple optical-depth estimate gives $N_{\rm SN}\tau\alpha\simeq 3$ at the DES upper limit, suggesting that the constraint lies near the Poissonian rare-event regime. In the ideal limit of negligible contamination, it would then improve approximately as $(N_{\rm SN}\tau)^{-1}$. The larger characteristic redshift of the LSST sample also increases the optical depth per supernova by a factor of order four relative to the DES sample. In practice, however, photometric misclassification, intrinsically bright supernova outliers, and uncertainty in the unlensed magnitude distribution introduce a per-supernova background and can restore an approximately $N_{\rm SN}^{-1/2}$ scaling. Moreover, constraining object fractions at the level of $\alpha\lesssim10^{-2}$ would require measuring and subtracting the foreground from stars and stellar remnants. We therefore retain $f_{\rm DM}\simeq0.018$ as a conservative fiducial projection in the point-like limit. 

Fig.~\ref{fig:edo_constraints} shows the resulting limits from DES and the projected LSST sensitivity, assuming a monochromatic EDO mass function. Compact objects with $r_{90}\ll R_E$ recover the point-lens bound, whereas the constraint weakens rapidly once the mass is distributed over scales larger than the Einstein radius.
We have added the resulting constraints for the current profile ($\rho \propto r^{-3/2}$), boson stars, and $\rho \propto r^{-9/4}$ to the \texttt{EDObounds} repository \cite{Croon:2024jhd,Sevi2024EDObounds} where they can also be conveniently recast for different assumptions for the mass function.

\section{Probing the primordial power spectrum}\label{sec:powerspectrum}

The map from a constraint on the UCMH abundance to a bound on the primordial curvature power spectrum proceeds through a sequence of modelling choices, each of which we make conservatively. 
Following Ref.~\cite{Bringmann:2025cht},  we present our constraints in terms of a power spectrum that is {\it locally scale-invariant},  $\mathcal{P}_\mathcal{R}(k) = \mathcal{P}_\mathcal{R}(k_\mathrm{R})$ for $k\sim k_R$.
 %
This allows for easy comparison to the literature, but makes shape-dependent comparisons with $y$/$\mu$-distortion limits somewhat model-dependent~\cite{Bringmann:2011ut}.

The further assumptions are:
(i) Press-Schechter collapse formalism; (ii) Gaussian statistics for $\delta_\chi$; (iii) a fixed, latest collapse redshift $z_c$, chosen conservatively, with later collapse yielding stronger but less robust limits; and (iv) the density profile assumed {(here $\rho \propto r^{-3/2}$)}, to which gravitational bounds are typically only weakly sensitive.\footnote{Note that constraints from DM annihilation (e.g. \cite{Bringmann:2011ut}) depend much more strongly on this assumption than the gravitational probe we consider here \cite{Delos:2018ueo}.}
We refer the reader to Refs.~\cite{Bringmann:2025cht,Bringmann:2011ut} for further details on the calculation. 

\begin{figure*}
    \centering
    \includegraphics[width=\linewidth]{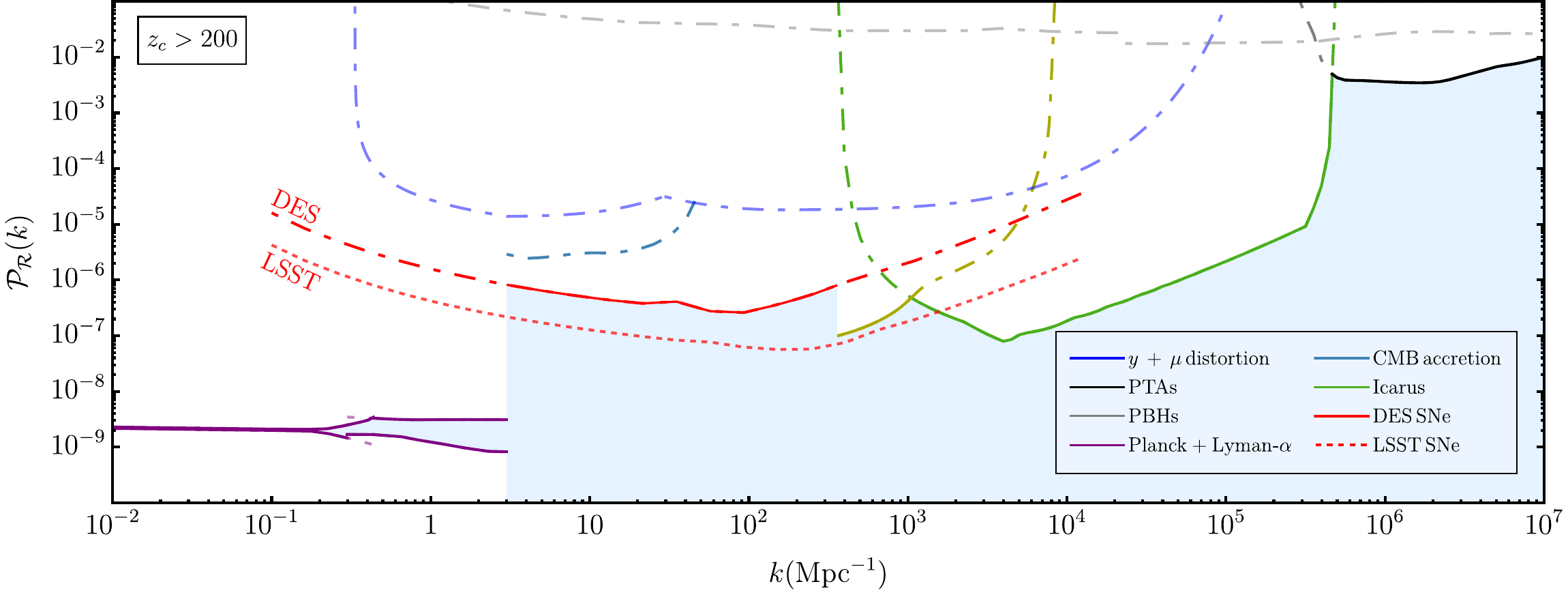}
    \caption{Constraints on the primordial power spectrum from DES \cite{DES:2024upw,DES:2024ffp} (continuous lines) and projections with LSST (dashed lines) calculated with the Press-Schechter formalism assuming collapse before $z_c = 200$. See text for further detail. 
    We also show constraints at large scales (purple) deriving from CMB and Lyman-$\alpha$  observations~\cite{Planck:2018vyg,Chluba:2015bqa,Green:2020jor}, 
    PBH bounds~\cite{Carr:2020gox},
    constraints from accretion \cite{Croon:2024rmw,Bringmann:2025cht},
    caustic crossings from Icarus~\cite{Bringmann:2025cht,Croon:2025yfj}, wide-binary evaporation~\cite{Olea-Romacho:2026pgn},
    $y$ and $\mu$ spectral distortions \cite{Chluba:2015bqa,Fixsen:1996nj}, and scalar-induced gravitational waves constrained by \cite{NANOGrav:2023hvm}.
    }
    \label{fig:powerspectrum}
\end{figure*}

In Fig.~\ref{fig:powerspectrum} we show the resulting limits on the primordial power spectrum, based on the UCMH constraints and projections in the previous section. These bounds are competitive over the range $3 \lesssim k/{\rm Mpc}^{-1} \lesssim 3\times10^2$, where existing constraints are dominated by CMB spectral distortions and, at smaller scales, by limits associated with accretion onto UCMH. The sensitivity of SN microlensing in this regime follows in part from its ability to probe perturbations that become nonlinear substantially after recombination: here we require collapse only by $z_c=200$, rather than restricting to structures already present at $z\gtrsim1100$. This lower collapse redshift reduces the primordial overdensity required for halo formation and therefore allows comparatively small enhancements of the primordial power spectrum to be constrained.

The change in sensitivity around $k\sim 10^2\,{\rm Mpc}^{-1}$ arises from the finite compactness required for the microlensing constraints. At smaller physical scales, perturbations that collapse at lower redshift form more extended halos, which eventually become too diffuse to produce an observable lensing signal; the constraint therefore becomes increasingly restricted to objects that formed at higher redshift. Likewise,
the derived constraints are only sensitive to the treatment of the optical-depth weighted efficiency Eq.~\ref{eq: alpha eps} at the large $k$ end; for smaller $z_S$ the Einstein radius becomes smaller, and sensitivity to more diffuse minihalos formed at lower redshifts is lost. We confirm numerically that $10 \%$ variations in $z_S$ lead to variations in the derived constraints at a similar level for $k \gtrsim 10^2 \rm Mpc^{-1}$, with vanishing variation below this scale. This dependence is therefore subdominant compared with the residual theoretical uncertainty associated with the modelling choices entering the UCMH abundance and structure, and primarily in the regime where other constraints dominate.

Two considerations limit the smallest scales to which SN microlensing is sensitive. First, finite-source effects become important when the lens mass falls below $M_{\rm ps}$, given in Eq.~\eqref{eq:Mfinitesource}. Since $M_{\rm UCMH}(k,z_{\rm eq})\propto k^{-3}$, the Einstein radius scales as $R_E\propto k^{-3/2}$, and finite-source suppression therefore turns on rapidly above a characteristic wavenumber $k_{\rm fs}$. Equating the UCMH mass associated with a mode $k$, evaluated at matter--radiation equality, to $M_{\rm ps}$ gives $k_{\rm fs}\sim {\rm few}\times10^4\,{\rm Mpc}^{-1}$, with a mild dependence on the lens and source distances and on the assumed photospheric radius.
Second, the point-lens constraint of Shah et al.\ is quoted only for masses $M>0.03\,M_\odot$. Using the same relation between UCMH mass and wavenumber as in our numerical calculation, the corresponding maximum wavenumber is defined by
\begin{equation}
    M_{\rm UCMH}
    \left(k_{\rm max}^{\rm DES},z_{\rm eq}\right)
    =
    0.03\,M_\odot,
\end{equation}
which gives $ k_{\rm max}^{\rm DES} \simeq 2\times10^4\,{\rm Mpc}^{-1}$.
As this is slightly more restrictive than our finite-source cutoff, we truncate both the DES recast and the LSST projection at $k_{\rm max}^{\rm DES}$. As shown in Fig.~\ref{fig:powerspectrum}, this restriction does not affect the range over which SN microlensing provides a competitive constraint.

\section{Discussion and outlook}\label{sec:conclusion}
In this work we have demonstrated that supernova microlensing probes dark matter structures formed from the collapse of primordial overdensities at small scales. We constructed an extended lens efficiency by comparing the EDO lens potential to that of a point-like lens, and used it to generalise the constraints found using the DES supernova sample \cite{DES:2024ffp}. Then, we demonstrated the implications for the primordial power spectrum of curvature perturbations under the assumptions discussed in Sec.~\ref{sec:powerspectrum}.

In this work we have not required that the small-scale power spectrum is predicted by, or correlated with, its amplitude on CMB scales. In this way our treatment differs from that of Ref.~\cite{DES:2024ffp}.
Here, following standard practice in primordial power spectrum constraints, we have treated the curvature power spectrum as \emph{locally} scale-invariant with that amplitude left free to be constrained rather than extrapolated from large scales. Our limits should therefore be read as constraints on the local amplitude of the small-scale power spectrum at the corresponding wavenumber, independent of the large-scale normalisation.

The Vera C. Rubin Observatory Legacy Survey of Space and Time will provide a qualitatively new opportunity to exploit this probe. The expected SN sample will be vastly larger than existing samples, making it possible to search statistically for lensing-induced magnification tails, anomalous light-curve perturbations, or excess scatter in standardized SN distances. This increase in power will also require careful control of systematics, such as supernovae outliers and distinguishing the effects of enhanced primordial fluctuations from standard CDM structures. 
{Simulation-based inference provides a promising framework for this task~\cite{Popovic:2026cxj}, allowing the intrinsic SN~Ia population and survey selection effects to be forward-modelled jointly with the microlensing signal and marginalised over when inferring the EDO abundance.}
Other refinements can also improve these limits, such as employing density-field reconstruction to delens the contribution from large halos~\cite{Porqueres:2021clw} or including EDOs in the analysis of strongly-lensed SNe~\cite{Suyu:2023jue}. Even in the absence of individual detections, such a sample can place projected limits on the abundance of EDOs over cosmological volumes.

\section*{Software}
Wolfram Mathematica version 14.2.1, python 3.13.7.

\section*{Acknowledgements}
The authors thank Torsten Bringmann, Maria Olalla Olea-Romacho, and Paul Shah for useful discussions. DC is supported by the STFC under Grant No.~ST/T001011/1 and ST/X003167/1. SSM is supported by funds provided by the Center for Particle Cosmology at the University of Pennsylvania.
MZ acknowledges support from grant RYC2024-049805-I, funded by MICIU/AEI/10.13039/501100011033 and by the European Social Fund Plus (ESF+), and from the Ayudas de Excelencia RYC-MaX 2024 programme of the Spanish National Research Council (CSIC). This publication has been funded within the framework of the R\&D\&I Project CEX2025-001574-S, funded by MICIU/AEI/10.13039/501100011033. The research presented in this publication falls within the research line Origin and Composition of the Universe: Astroparticles and Cosmology (Astro/Cosmo). This work is supported by ERC grant GLOW (101230608). Funded by the European Union. Views and opinions expressed are however those of the author(s) only and do not necessarily reflect those of the European Union or the European Research Council Executive Agency. Neither the European Union nor the granting authority can be held responsible for them.

\bibliography{refs}

\end{document}

%% file: universalnewcommands.tex
\renewcommand{\tilde}{\widetilde} 

\newcommand{\beq}{\begin{equation}}
\newcommand{\eeq}{\end{equation}}
\newcommand{\bea}{\begin{eqnarray}}
\newcommand{\eea}{\end{eqnarray}}

\usepackage{pdfbase}[2017/03/16]
\usepackage{xparse,ocgbase}
\usepackage{xcolor,calc}
\usepackage{tikzpagenodes,linegoal}
\usetikzlibrary{calc}
\usepackage{tcolorbox}

\ExplSyntaxOn
\let\tpPdfLink\pbs_pdflink:nn
\let\tpPdfAnnot\pbs_pdfannot:nnnn\let\tpPdfLastAnn\pbs_pdflastann:
\let\tpAppendToFields\pbs_appendtofields:n
\def\tpPdfXform{\pbs_pdfxform:nnnnn{1}{1}{}{}}
\let\tpPdfLastXform\pbs_pdflastxform:
\let\cListSet\clist_set:Nn\let\cListItem\clist_item:Nn
\ExplSyntaxOff

\usepackage{pdfbase}[2017/03/16]
\usepackage{xparse,ocgbase}
\usepackage{xcolor,calc}
\usepackage{tikzpagenodes,linegoal}
\usetikzlibrary{calc}
\usepackage{tcolorbox}

\ExplSyntaxOn
\let\tpPdfLink\pbs_pdflink:nn
\let\tpPdfAnnot\pbs_pdfannot:nnnn\let\tpPdfLastAnn\pbs_pdflastann:
\let\tpAppendToFields\pbs_appendtofields:n
\def\tpPdfXform{\pbs_pdfxform:nnnnn{1}{1}{}{}}
\let\tpPdfLastXform\pbs_pdflastxform:
\let\cListSet\clist_set:Nn\let\cListItem\clist_item:Nn
\ExplSyntaxOff

\makeatletter
\NewDocumentCommand{\tooltip}{%
  ssssO{\ifdefined\@linkcolor\@linkcolor\else blue\fi}mO{yellow!20}mO{0pt,0pt}%
}{{%
  \leavevmode%
  \IfBooleanT{#2}{%
    \ocgbase@new@ocg{tipOCG.\thetcnt}{%
      /Print<</PrintState/OFF>>/Export<</ExportState/OFF>>%
    }{false}%
    \xdef\tpTipOcg{\ocgbase@last@ocg}%
    \ocgbase@add@ocg@to@radiobtn@grp{tool@tips}{\ocgbase@last@ocg}%
  }%
  \tpPdfLink{%
    \IfBooleanTF{#4}{%
      /Subtype/Link/Border[0 0 0]/A <</S/SetOCGState/State [/Toggle \tpTipOcg]>>
    }{%
      /Subtype/Screen%
      /AA<<%
        \IfBooleanTF{#3}{%
          /E<</S/SetOCGState/State [/Toggle \tpTipOcg]>>%
        }{%
          \IfBooleanTF{#2}{%
            /E<</S/SetOCGState/State [/ON \tpTipOcg]>>%
            /X<</S/SetOCGState/State [/OFF \tpTipOcg]>>%
          }{
            \IfBooleanTF{#1}{%
              /E<</S/JavaScript/JS(%
                var fd=this.getField('tip.\thetcnt');%
                if(typeof(click\thetcnt)=='undefined'){%
                  var click\thetcnt=false;%
                  var fdor\thetcnt=fd.rect;var dragging\thetcnt=false;%
                }%
                if(fd.display==display.hidden){%
                  fd.delay=true;fd.display=display.visible;fd.delay=false;%
                }else{%
                  if(!click\thetcnt&&!dragging\thetcnt){fd.display=display.hidden;}%
                  if(!dragging\thetcnt){click\thetcnt=false;}%
                }%
                this.dirty=false;%
              )>>%
            }{%
              /E<</S/JavaScript/JS(%
                var fd=this.getField('tip.\thetcnt');%
                if(typeof(click\thetcnt)=='undefined'){%
                  var click\thetcnt=false;%
                  var fdor\thetcnt=fd.rect;var dragging\thetcnt=false;%
                }%
                if(fd.display==display.hidden){%
                  fd.delay=true;fd.display=display.visible;fd.delay=false;%
                }%
               this.dirty=false;%
              )>>%
              /X<</S/JavaScript/JS(%
                if(!click\thetcnt&&!dragging\thetcnt){fd.display=display.hidden;}%
                if(!dragging\thetcnt){click\thetcnt=false;}%
                this.dirty=false;%
              )>>%
            }%
            /U<</S/JavaScript/JS(click\thetcnt=true;this.dirty=false;)>>%
            /PC<</S/JavaScript/JS (%
              var fd=this.getField('tip.\thetcnt');%
              try{fd.rect=fdor\thetcnt;}catch(e){}%
              fd.display=display.hidden;this.dirty=false;%
            )>>%
            /PO<</S/JavaScript/JS(this.dirty=false;)>>%
          }%
        }%
      >>%
    }%
  }{{\color{#5}#6}}%
  \sbox\tiptext{%
    \IfBooleanT{#2}{%
      \ocgbase@oc@bdc{\tpTipOcg}\ocgbase@open@stack@push{\tpTipOcg}}%
    \tcbox[colframe=black,colback=#7,size=fbox,arc=1ex,sharp corners=southwest]{#8}%
    \IfBooleanT{#2}{\ocgbase@oc@emc\ocgbase@open@stack@pop\tpNull}%
  }%
  \cListSet\tpOffsets{#9}%
  \edef\twd{\the\wd\tiptext}%
  \edef\tht{\the\ht\tiptext}%
  \edef\tdp{\the\dp\tiptext}%
  \tipshift=0pt%
  \IfBooleanTF{#2}{%
    \setlength\whatsleft{\linegoal}%
  }{%
    \measureremainder{\whatsleft}%
  }%
  \ifdim\whatsleft<\dimexpr\twd+\cListItem\tpOffsets{1}\relax%
    \setlength\tipshift{\whatsleft-\twd-\cListItem\tpOffsets{1}}\fi%
  \IfBooleanF{#2}{\tpPdfXform{\tiptext}}%
  \raisebox{\heightof{#6}+\tdp+\cListItem\tpOffsets{2}}[0pt][0pt]{%
    \makebox[0pt][l]{\hspace{\dimexpr\tipshift+\cListItem\tpOffsets{1}\relax}%
    \IfBooleanTF{#2}{\usebox{\tiptext}}{%
      \tpPdfAnnot{\twd}{\tht}{\tdp}{%
        /Subtype/Widget/FT/Btn/T (tip.\thetcnt)%
        /AP<</N \tpPdfLastXform>>%
        /MK<</TP 1/I \tpPdfLastXform/IF<</S/A/FB true/A [0.0 0.0]>>>>%
        /Ff 65536/F 3%
        /AA <<%
          /U <<%
            /S/JavaScript/JS(%
              var fd=event.target;%
              var mX=this.mouseX;var mY=this.mouseY;%
              var drag=function(){%
                var nX=this.mouseX;var nY=this.mouseY;%
                var dX=nX-mX;var dY=nY-mY;%
                var fdr=fd.rect;%
                fdr[0]+=dX;fdr[1]+=dY;fdr[2]+=dX;fdr[3]+=dY;%
                fd.rect=fdr;mX=nX;mY=nY;%
              };%
              if(!dragging\thetcnt){%
                dragging\thetcnt=true;Int=app.setInterval("drag()",1);%
              }%
              else{app.clearInterval(Int);dragging\thetcnt=false;}%
              this.dirty=false;%
            )%
          >>%
        >>%
      }%
      \tpAppendToFields{\tpPdfLastAnn}%
    }%
  }}%
  \stepcounter{tcnt}%
}}
\makeatother
\newsavebox\tiptext\newcounter{tcnt}
\newlength{\whatsleft}\newlength{\tipshift}
\newcommand{\measureremainder}[1]{%
  \begin{tikzpicture}[overlay,remember picture]
    \path let \p0 = (0,0), \p1 = (current page.east) in
      [/utils/exec={\pgfmathsetlength#1{\x1-\x0}\global#1=#1}];
  \end{tikzpicture}%
}